# Ultrafast correlated charge and lattice motion in a hybrid metal halide perovskite

**Authors:** Yang Lan[1], Benjamin J. Dringoli[1], David A. Valverde-Chavez[1], Carlito S. Ponseca Jr.[2], Mark Sutton[1], Yihui He[3], Mercouri G. Kanatzidis[3], & David G. Cooke[1]*

**Affiliations:**

[1]Department of Physics, McGill University, Montreal, QC H3A 2T8, Canada.

[2]Division of Biomolecular and Organic Electronics, IFM, Linköpings Universitet, Linköping, SE- 58183, Sweden.

[3]Department of Chemistry, Northwestern University, Evanston, Illinois 60208, USA.

*Correspondence to: cooke@physics.mcgill.ca

**Abstract:**

Hybrid organic-inorganic halide perovskites have shown remarkable optoelectronic properties (*1-3*), believed to originate from correlated motion of charge carriers and the polar lattice forming large polarons (*4-7*). Few experimental techniques are capable of probing these correlations directly, requiring simultaneous sub-meV energy and femtosecond temporal resolution after absorption of a photon (*8*). Here we use transient multi-THz spectroscopy, sensitive to the internal motions of charges within the polaron, to temporally and energetically resolve the coherent coupling of charges to longitudinal optical phonons in single crystal $CH_3NH_3PbI_3$ (MAPI). We observe room temperature quantum beats arising from the coherent displacement of charge from the coupled phonon cloud. Our measurements provide unambiguous evidence of the existence of polarons in MAPI.

**One Sentence Summary:**

Polarons in hybrid lead halide perovskites are observed by ultrafast terahertz spectroscopy showing coherent, correlated motion of charge carriers with coupled lattice vibrations.

**Main Text:**

Electron-phonon coupling appears to play a significant role in defining the optoelectronic properties of hybrid organic-inorganic halide perovskites (*7, 9*). The perovskite lattice, with chemical formula $ABX_3$, is composed of an octahedrally coordinated metal-halide $BX_3$ sub-lattice and an organic A-site cation that displays dynamic disorder due to 'soft' non-covalent bonding. This results in a complex vibrational landscape in a highly polar material. A charge moving in such a polar lattice causes a local distortion as atoms move in response to the Coulomb perturbation (*8, 10, 11*). This charge and coupled lattice distortion move as a single quasiparticle called a polaron (*12-14*). Dynamic screening of electron-hole Coulomb interactions by the motion of the polar lattice can lower scattering rates, which may explain the significant carrier mobilities in defect-heavy solution processed films and long carrier lifetimes beyond that of a simple direct band gap semiconductor. Ultrafast Kerr, Raman and 2D electronic spectroscopy measurements have shown coherent nuclear vibrational dynamics impulsively coupled to electronic inter-band transitions (*5, 6, 15-19*). Electronic coherence, however, is typically not observed, dominated instead by a solvation-like response suggesting dynamic screening by the freely rotating organic cation (*6*) or from the anharmonic response of the perovskite structure itself (*15*). THz spectroscopy, which probes intraband carrier motion, has shown a delayed onset of photoconductivity, taking several hundreds of femtoseconds to reach a maximum. This was initially interpreted as a signature of exciton dissociation (*20, 21*), however more recently it has been assigned to the dressing of charge carriers by LO phonons, forming a large polaron which while increasing the carrier effective mass also reduces momentum scattering (*22*). THz emission from perovskite films also show coherent vibrational dynamics at a lower LO phonon energy of 1.2 THz (*23*). There is significant uncertainty in the nature of polaron formation in these perovskite materials, as well as the role each phonon plays in its stabilization (*24*).

In this work we use time-resolved multi-THz spectroscopy (TRTS) to probe the correlated motion of charges in single crystal MAPI at room temperature. A schematic of the experiment is shown in Fig. 1A, where an optical pump pulse resonant with the band edge injects charge carriers, and the intraband polarization is probed by reflection of a single-cycle, ultrabroadband THz pulse at delay time τ after photoexcitation. Coherent detection of the THz pulse electric field allows the resolution of each Fourier component within the pulse (amplitude and phase) with sub-meV energy resolution, and the dynamics of their modulation followed with 35 fs temporal resolution. We use this capability to resolve the coherent motion of charge carriers correlated to the motion of the polar lattice of MAPI, as polarons are formed on a time scale commensurate with the LO phonon period after band edge excitation.

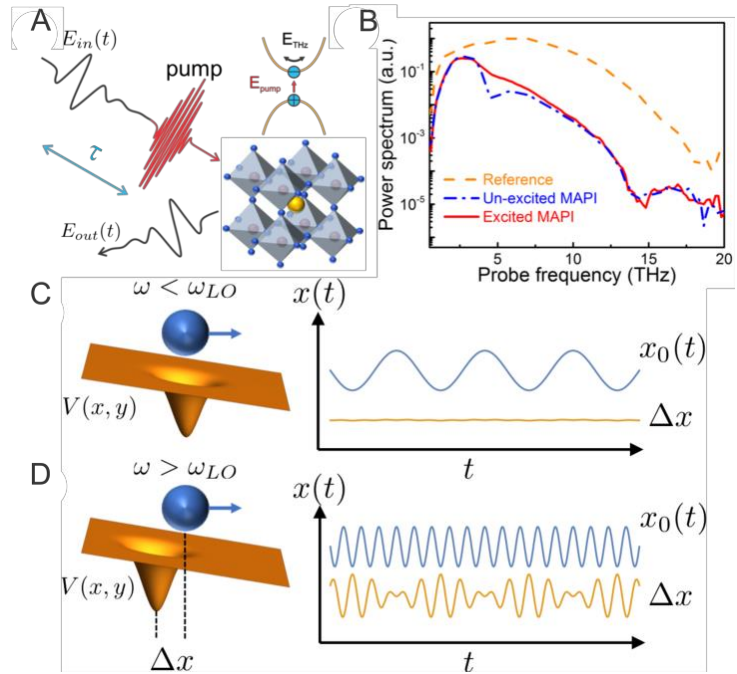

**Fig. 1. Schematic of TRTS measurement and analogy of a harmonically driven oscillator.** (**A**) Schematic of TRTS on a single-crystal MAPI in reflection configuration with pump-probe delay time τ. (**B**) Reflected THz power spectrum from a metallic mirror (orange dash line), unexcited MAPI (blue dash dot curve), and MAPI photoexcited at τ = 1 ps (red solid line). Schematic of the internal motion of a driven charge at frequency ω coupled to a LO polar lattice distortion with phonon frequency $\omega_{LO}$ for frequencies above (**C**) and below (**D**) resonance, with the accompanying motion of the bare charge, $x_0(t)$ and the relative displacement $\Delta x(t)$ showing a beat.

THz spectroscopy is a probe of intraband polarization, governed by charge centre-of-mass motion as well as the internal displacement of charge within the quasiparticle (*25*), schematically depicted in Figs. 1(C, D). The intraband polarization arising from internal charge displacement is analogous to a harmonically driven oscillator, e.g. a pendulum. The relative displacement for a driven nonlinear harmonic oscillator will display beats between the driving frequency ω and the characteristic frequency of the oscillator. For the polaron, the local THz electric field drives charge displacement at frequency ω, and the heavier lattice distortion follows at a characteristic frequency of the LO phonon at $\omega_{LO}$. For $\omega \ll \omega_{LO}$ only centre-of-mass motion is possible and the lattice distortion follows the charge carrier, so no beat is observed in their relative motion. Near resonance, $\omega \approx \omega_{LO}$, a small oscillation is expected for a frequency range defined by the uncertainty relation and the lifetime of the LO phonon. A π phase flip in the relative displacement is expected when ω is further increased and crosses resonance. For $\omega \gg \omega_{LO}$, a quantum beat between the intraband motion of free charges and the coupled lattice distortion is expected. The intraband polarization is subsequently modulated at the beat frequency, which is detected by THz reflectivity. The 35 fs optical excitation impulsively launches phase coherent motion for probe energies below ~10 THz, and scanning the pump-probe delay $\tau$ maps the time evolution of this coherent beat until it dephases due to interactions with the thermal bath.

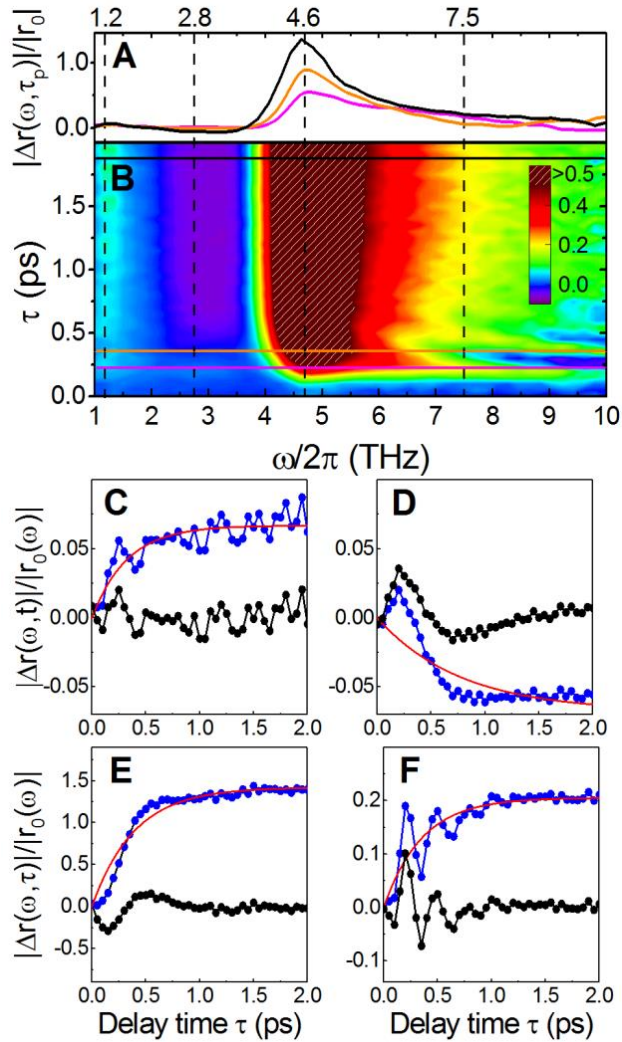

**Fig. 2. Time-resolved THz reflectivity from MAPI after band-edge optical excitation.** (**A**) Differential reflectivity $|\Delta r(\omega,\tau)|/|r_0|$ at selected delay time τ marked in (**B**). (**B**) Two-dimensional probe frequency/delay time color scale map for $|\Delta r(\omega,\tau)|/|r_0|$. (**C-F**) $|\Delta r(\omega,\tau)|/|r_0|$ at the four probe frequencies 1.2 THz, 2.8 THz, 4.6 THz, and 7.5 THz, respectively. The differential reflectivity amplitude is represented by blue curves, the fitted background exponential change is shown by red curves, and the differential reflectivity oscillations obtained by subtracting the slow exponential rise are shown in black.

The reflected THz pulse power spectrum from the crystal at negative pump-probe delays, before the pump pulse has arrived, is shown in Fig. 1B along with the $\tau = 1$ ps spectrum. As in previous work (*21*), the un-pumped power spectrum is dominated by a dip at $\omega \sim (2\pi)$ 4.8 THz due to a reflectivity minimum on the high frequency edge of the Restrahlen band, which occurs between the TO phonon at $\omega_{TO} = (2\pi)$ 2 THz and the LO phonon at $\omega_{LO} = (2\pi)$ 3.7 THz (*21, 26, 27*). Upon excitation, mobile charge carriers move to screen the polar lattice, eliminating the 4.8 THz dip in the reflected THz pulse as shown in the red line in Fig. 1B. The amplitude of the differential THz reflectivity, $|\Delta r(\omega,\tau)|/|r_0|$ is therefore strongly peaked at the nearby frequency of 4.6 THz, as shown in the fixed $\tau$ slices of the 2D map in Fig. 2A.

Close inspection of the early time dynamics in the $|\Delta r(\omega,\tau)|/|r_0|$ map in Fig. 2B shows coherent oscillations appearing along the pump-probe delay axis for frequencies higher than the main peak, superimposed on an approximately exponential, sub-picosecond rise to the steady state. These oscillations reflect the internal motion of the charge within the polaron. One dimensional delay time cuts for fixed ω are shown in Figs. 2(C-F) for probe energies indicated in Fig. 2B. At the lowest frequency of $\omega = (2\pi)$ 1.2 THz in Fig. 2C, just above the Pb-I related TO phonon at 1 THz (*26, 27*) and in a region where the contribution of the polar lattice dominates, the differential reflectivity is small as the injected charge is a weak perturbation relative to the reflectivity of the unexcited lattice. No oscillations are present in this case within the limit of scan time. At $\omega = (2\pi)$ 2.9 THz (12 meV) in Fig. 2D the sign of the steady state differential reflectivity is initially positive but rapidly turns negative within 250 fs. A negative $|\Delta r|/|r_0|$ is expected in this region as mobile carriers move to screen lattice motion, reducing the reflectivity relative to the unexcited state. For Fig. 2E, $\omega = (2\pi)$ 4.6 THz where the lattice is least polarizable, small oscillations are superimposed on the monotonically increasing signal. At $\omega = (2\pi)$ 7.5 THz in Fig. 2F, electronic contributions dominate and high frequency oscillations are observed lasting for several periods.

The slow contributions to the signal can be well described by an exponential rise convoluted with a Gaussian instrument response function of 35 fs and centred at $\tau = 0$. Fits to the selected cuts described above are shown in Figs. 2(C-F), which in all cases are well described by a sub-picosecond rise time to be discussed later. Subtraction of this slow response reveals the isolated coherent oscillations, $|\Delta r(\omega,\tau)|_{osc}$, that rapidly decay on a sub-picosecond time scale commensurate with the slow rise. Fig. 3A shows the pure oscillatory component of the normalized $|\Delta r(\omega,\tau)|_{osc}^{norm}$ for all THz probe energies. For $\omega < \omega_{LO}$, a ~ 5 times weaker, highly damped oscillation at a fixed frequency of ~ 0.5 THz is observed, which is the low frequency limit of our scans. The sign of the oscillations undergoes a $\pi$ phase shift at $\omega = \omega_{LO}$, as expected when transiting the resonance. For $\omega > \omega_{LO}$ the oscillations gradually shift to higher frequencies with no further change in phase. The amplitude of the coherent oscillation is shown in Fig. 4A, showing a marked turn on at $\omega_{LO}$ and rising to a peak at 4.6 THz, corresponding to the spectral region where the lattice is least polarizable just above the Restrahlen band. The full width at half-maximum of this peak is approximately 1 THz, comparable to the lifetime of the LO phonon at room temperature (*27*), broadened due to mode coupling with the highly damped and disordered motion of the organic cation.

The interaction of charge carriers with optical phonons is governed by a dressed Coulomb interaction, $W_q = \frac{V_q}{\varepsilon(\omega,\tau)}$, modified through dynamic screening by the correlated lattice dynamics described by the lattice dielectric function ε(ω,τ). The dielectric loss function, $-Im[1/\tilde{\varepsilon}(\omega,\tau)]$, schematically plotted in Fig. 3B following Ref. (*28*), describes energy loss through interaction with optical phonons and is centered at ω$_{LO}$, with a very broad ~1 THz FWHM due to mode coupling with the organic cation (*27*). The screening function, $Re[1/\tilde{\varepsilon}(\omega,\tau)]$ renormalizes the charge in the dressed Coulomb interaction experienced by another charge under energy exchange of ħω, and proceeds from an over-screened response at ω < ω$_{LO}$, where the lattice efficiently moves to oppose Coulomb forces between charge carriers, to ω > ω$_{LO}$ where the lattice cannot follow the electromagnetic perturbation and under-screens the interaction. The sign change in the coherent oscillation observed in Fig. 3A corresponds to the transition from an under-screened to an over-screened lattice response, where the lattice polarization moves from an in-phase to an out-of-phase response with the electromagnetic perturbation (*29*).

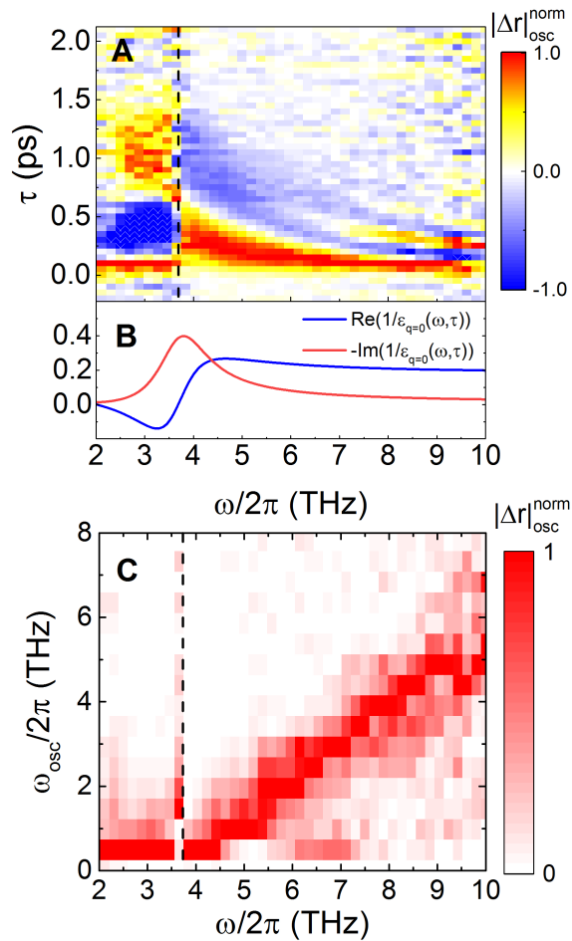

**Fig. 3. Coherent oscillations of THz reflectivity without slow rise component.** (**A**) Normalized $|\Delta r(\omega,\tau)|_{osc}^{norm}$ map with slow rise component subtracted. The vertical dash line marks the LO phonon frequency at 3.7 THz where a π phase flip occurs and above which a subsequent frequency chirp is evident. (**B**) Schematic dielectric energy loss function describing inelastic optical phonon scattering and screening function describing the renormalization of dressed charges for MAPI. (**C**) Fourier transformation along delay time axis τ, $|\Delta r(\omega,\omega_{osc})|_{osc}^{norm}$, showing a clear onset of a coherent beat between charge motion at ω and ω$_{LO}$.

A Fourier transform of this normalized differential reflectivity along the pump-probe delay axis yields a two-dimensional $|\Delta r(\omega, \omega_{osc})|_{osc}$ map shown in Fig. 3C. The map shows a clear coherent oscillation frequency $\omega_{osc} \propto \omega - \omega_{LO}$, or the beat frequency of the charge motion coupled to a lattice distortion. The coherent oscillations are governed by a single LO phonon at $\omega_{LO} = (2\pi)$ 3.7 THz, 15 meV or 125 cm$^{-1}$ as shown by the $\omega$-intercept at $\omega_{osc} = 0$ and the location of the phase flip in Fig. 3A. This is in agreement with time-domain THz reflectivity measurements on a MAPI single crystal showing a single effective LO phonon at ~3.9 THz, which was temperature independent and strongly damped due to coupling with the dynamic orientational motion of the MA cation (*27*). Another recent *ab initio* calculation aimed at understanding the role each phonon plays in polaron correlations in MAPI showed multiple phonon branches contribute: bending (4 meV) and stretching (13 meV) modes of the Pb-I sublattice and translational+librational modes of the organic cation (20 meV). At room temperature the considerable broadening of the latter two lines merges to one single effective LO phonon, in excellent agreement with the observed $\hbar\omega = 15$ meV (*24*).

The Fröhlich electron-phonon coupling constant α can subsequently be calculated by (*14*)

$$\alpha = \frac{e^2}{\hbar \omega_{LO}} \frac{1}{4\pi\varepsilon_0} \left(\frac{m^*}{2\hbar\omega_{LO}}\right)^{\frac{1}{2}} \left(\frac{1}{\varepsilon_s} - \frac{1}{\varepsilon_\infty}\right) \quad (1)$$

where $m^*$ is the effective mass, $\varepsilon_s$ and $\varepsilon_\infty$ are the dielectric constants below and above the related Restrahlen band, with parameters given in the Supplementary Materials. This yields a value of

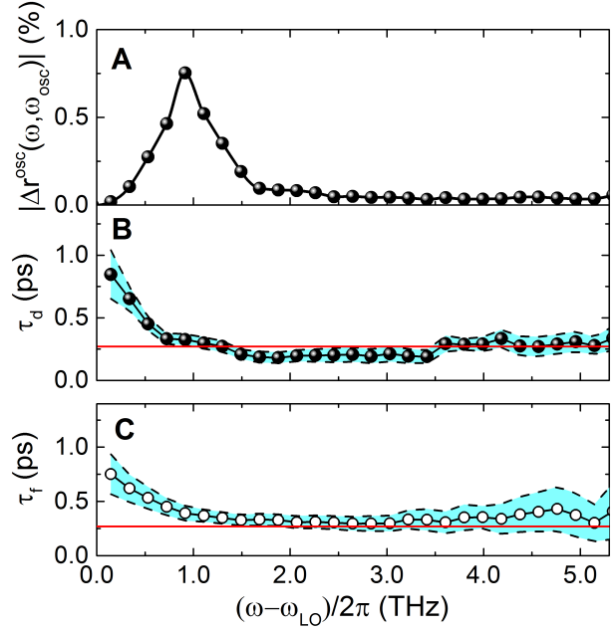

**Fig. 4. Amplitude and time constants.** (**A**) Amplitude of oscillating differential THz reflectivity relative to the total differential reflectivity. (**B**) The fitted dephasing time of the coherent oscillations and (**C**) the polaron formation time using the model discussed in the text. Shaded regions indicate 95% confidence regimes in the fitting. The red line in (**B**) and (**C**) marks one LO phonon period.

α =2.0, from which a polaron binding energy can be calculated by $E_P = -\hbar\omega_{LO}[\alpha + 0.0158\alpha^2 + 0.00081\alpha^3] \approx 30$ meV and the radius of the polaron can be estimated by $r_p = \sqrt{\frac{\hbar}{2m^*\omega_{LO}}} \approx 3.5$ nm or approximately 6 unit cells. This places MAPI in the intermediate coupling regime similar to other metal halides.

The dynamics of $|\Delta r(\tau)|/|r_0|$ is described by fitting the data to a slow exponential rise $R_1(\tau) = \theta(\tau)(a + be^{-\tau/\tau_f})$ and a damped oscillator $R_2(\tau) = \theta(\tau)c\, e^{-\tau/\tau_d}\cos[(\omega - \omega_{LO})\tau + \varphi]$. The $R_1(\tau)$ component models the dynamics of polaron formation with a time constant $\tau_f$, and $R_2(\tau)$ describes the quantum beat dephasing with a time constant $\tau_d$. The results of these fits are shown in Figs. 4(B, C), with both time constants comparable and increasing near the resonance condition ω = ω<sub>LO</sub> and relatively constant at 200 - 300 fs above 5 THz. This corresponds to approximately one LO phonon period (270 fs), indicating rapid dephasing potentially through mode coupling with the organic cation (*27*) or through lattice anharmonicity (*15*).

A fully-quantum, many-body description of polaron formation was formulated in Ref. (*30*) solving the time-dependent Schrödinger equation for the Holstein model of electron-phonon coupling after launching an initially free electron wave packet. While an injected electron interacts with the radiated LO phonons upon initial injection, the electron wave packet and all other correlation functions will oscillate between that of an initial free carrier state and a fully dressed polaron at a frequency given by the LO phonon. Since the polaron is mobile through its excited states, the probability of measuring a free particle $P(\tau) \propto \cos[(\omega - \omega_{LO})\tau]$ oscillates at the difference between the ground and the excited state continuum probed at ω. The beat only appears for ω> ω<sub>LO</sub> as this is the required energy to probe the excited states, above which there exists a quasi-continuum of states (*24*). The room temperature broadening of the LO phonon through mode coupling with the organic cation has put into question the role of polaron effects at room temperature (*27*). Our results show that polaron correlations are present almost instantaneously after charge injection and their coherent interaction with their radiated LO phonons decaying on a time scale comparable to the phonon period. This decay signifies the formation time of a stable polaron.

In summary, we report the measurement of intraband quantum beats of polarons in MAPI as they are formed on a sub-picosecond time scale at room temperature. The origin of the quantum beat can be understood as the relative motion of the charge carriers to their coupled LO phonon cloud. We find that a single effective LO phonon at ω<sub>LO</sub> = (2π) 3.7 THz is coupled to electronic motion, which has been identified as a mixture of two modes, a Pb-I stretch and a librational and transitional motion of the MA cation. The slow rise time connected to polaron formation and the dephasing time of the coherent oscillation are highly damped at approximately one phonon period. These measurements of correlated electron-phonon motion in the band provide direct and quantitative evidence that stable polarons exist in MAPI at room temperature. Direct experimental signatures of polarons in HOIPs have proven difficult until now, and first principles calculations are impeded by the large number of atoms involved. Confirmation of the polaronic nature of carriers in the lead halide perovskites will have significant consequences for the morphological and compositional engineering of perovskite devices, as it is clear that optimizing the electronic structure cannot be accomplished without accounting for their intricate, correlated dance with the lattice.

**References and Notes:**

1. C. C. Stoumpos, M. G. Kanatzidis, The Renaissance of Halide Perovskites and Their Evolution as Emerging Semiconductors. *Accounts Chem Res* **48**, 2791-2802 (2015).
2. S. D. Stranks, H. J. Snaith, Metal-halide perovskites for photovoltaic and light-emitting devices. *Nat Nanotechnol* **10**, 391-402 (2015).
3. M. A. Green, A. Ho-Baillie, Perovskite Solar Cells: The Birth of a New Era in Photovoltaics. *ACS Energy Letters* **2**, 822-830 (2017).
4. X. Y. Zhu, V. Podzorov, Charge Carriers in Hybrid Organic–Inorganic Lead Halide Perovskites Might Be Protected as Large Polarons. *The Journal of Physical Chemistry Letters* **6**, 4758-4761 (2015).
5. H. M. Zhu *et al.*, Screening in crystalline liquids protects energetic carriers in hybrid perovskites. *Science* **353**, 1409-1413 (2016).
6. K. Miyata *et al.*, Large polarons in lead halide perovskites. *Sci Adv* **3**, (2017).
7. M. Bonn, K. Miyata, E. Hendry, X. Y. Zhu, Role of Dielectric Drag in Polaron Mobility in Lead Halide Perovskites. *ACS Energy Letters* **2**, 2555-2562 (2017).
8. M. Betz *et al.*, Subthreshold carrier-LO phonon dynamics in semiconductors with intermediate polaron coupling: A purely quantum kinetic relaxation channel. *Physical Review Letters* **86**, 4684-4687 (2001).
9. K. Miyata, T. L. Atallah, X. Y. Zhu, Lead halide perovskites: Crystal-liquid duality, phonon glass electron crystals, and large polaron formation. *Sci Adv* **3**, (2017).
10. T. D. Lee, F. E. Low, D. Pines, The Motion of Slow Electrons in a Polar Crystal. *Physical Review* **90**, 297-302 (1953).
11. A. Leitenstorfer, S. Hunsche, J. Shah, M. C. Nuss, W. H. Knox, Femtosecond Charge Transport in Polar Semiconductors. *Physical Review Letters* **82**, 5140-5143 (1999).
12. H. Frölich, Electrons in lattice fields. *Advances in Physics* **3**, 325-361 (1954).
13. R. P. Feynman, Slow Electrons in a Polar Crystal. *Physical Review* **97**, 660-665 (1955).
14. J. T. Devreese, Polarons, AIP Conference Proceedings **678,** 2003.
15. O. Yaffe *et al.*, Local Polar Fluctuations in Lead Halide Perovskite Crystals. *Physical Review Letters* **118**, (2017).
16. D. M. Monahan *et al.*, Room-Temperature Coherent Optical Phonon in 2D Electronic Spectra of CH3NH3PbI3 Perovskite as a Possible Cooling Bottleneck. *J Phys Chem Lett* **8**, 3211-3215 (2017).
17. T. Ghosh, S. Aharon, L. Etgar, S. Ruhman, Free Carrier Emergence and Onset of Electron-Phonon Coupling in Methylammonium Lead Halide Perovskite Films. *J Am Chem Soc* **139**, 18262-18270 (2017).
18. G. Batignani *et al.*, Probing femtosecond lattice displacement upon photo-carrier generation in lead halide perovskite. *Nat Commun* **9**, 1971 (2018).
19. F. Thouin *et al.*, Phonon coherences reveal the polaronic character of excitons in two-dimensional lead-halide perovskites. *arXiv*, 1-22 (2018).
20. C. S. Ponseca *et al.*, Organometal Halide Perovskite Solar Cell Materials Rationalized: Ultrafast Charge Generation, High and Microsecond-Long Balanced Mobilities, and Slow Recombination. *Journal of the American Chemical Society* **136**, 5189-5192 (2014).

21. D. A. Valverde-Chávez *et al.*, Intrinsic femtosecond charge generation dynamics in single crystal CH3NH3PbI3. *Energ Environ Sci* **8**, 3700-3707 (2015).
22. S. A. Bretschneider *et al.*, Quantifying Polaron Formation and Charge Carrier Cooling in Lead-Iodide Perovskites. *Advanced Materials* **30**, (2018).
23. B. Guzelturk *et al.*, Terahertz Emission from Hybrid Perovskites Driven by Ultrafast Charge Separation and Strong Electron-Phonon Coupling. *Advanced Materials* **30**, (2018).
24. M. Schlipf, S. Poncé, F. Giustino, Carrier Lifetimes and Polaronic Mass Enhancement in the Hybrid Halide Perovskite CH3NH3PbI3 from Multiphonon Fröhlich Coupling. *Physical Review Letters* **121**, 086402 (2018).
25. P. Gaal *et al.*, Internal motions of a quasiparticle governing its ultrafast nonlinear response. *Nature* **450**, 1210-1213 (2007).
26. C. La-o-vorakiat *et al.*, Phonon Mode Transformation Across the Orthohombic-Tetragonal Phase Transition in a Lead Iodide Perovskite CH3NH3PbI3: A Terahertz Time-Domain Spectroscopy Approach. *Journal of Physical Chemistry Letters* **7**, 1-6 (2016).
27. M. Nagai *et al.*, Longitudinal Optical Phonons Modified by Organic Molecular Cation Motions in Organic-Inorganic Hybrid Perovskites. *Physical Review Letters* **121**, 145506 (2018).
28. L. M. Herz, How Lattice Dynamics Moderate the Electronic Properties of Metal-Halide Perovskites. *The Journal of Physical Chemistry Letters*, 6853-6863 (2018).
29. R. Huber *et al.*, How many-particle interactions develop after ultrafast excitation of an electron-hole plasma. *Nature* **414**, 286-289 (2001).
30. C. Kubler *et al.*, Coherent structural dynamics and electronic correlations during an ultrafast insulator-to-metal phase transition in VO2. *Phys Rev Lett* **99**, 116401 (2007).
31. He, Y. *et al.* Resolving the Energy of γ-Ray Photons with MAPbI3 Single Crystals. *ACS Photonics* **5**, 4132-4138 (2018).
32. G. Giorgi, J.-I. Fujisawa, H. Segawa, K. J. T. j. o. p. c. l. Yamashita, Small photocarrier effective masses featuring ambipolar transport in methylammonium lead iodide perovskite: a density functional analysis. *The journal of physical chemistry letters* **4**, 4213-4216 (2013).
33. M. Sendner *et al.*, Optical phonons in methylammonium lead halide perovskites and implications for charge transport. *Materials Horizons* **3**, 613-620 (2016).

**Acknowledgments:**

**Funding:**
Cooke and Sutton would like to acknowledge financial support from NSERC, CFI and FQRNT. Work at Northwestern (synthesis and fundamental studies of halide perovskites) was supported by U.S. DOE, Office of Science (grant SC0012541).

**Author contributions:**
*Conceptualization:* David G. Cooke
*Data curation:* Yang Lan
*Formal analysis:* Yang Lan, Benjamin J. Dringoli and David G. Cooke
*Funding acquisition:* Mark Sutton, Mercouri G. Kanatzidis and David G. Cooke

*Investigation:* Yang Lan, Benjamin J. Dringoli, Carlito S. Ponseca Jr. and David A. Valverde-Chavez
*Methodology:* Yang Lan and David G. Cooke
*Project administration:* David G. Cooke
*Resources:* Yihui He and Mercouri G. Kanatzidis
*Software:* Yang Lan, Benjamin J. Dringoli, David A. Valverde-Chavez and David G. Cooke
*Supervision:* Mark Sutton, Mercouri G. Kanatzidis and David G. Cooke
*Validation:* David G. Cooke
*Visualization:* Yang Lan and Benjamin J. Dringoli
*Writing – original draft:* Yang Lan
*Writing – review & editing:* Benjamin J. Dringoli, Carlito S. Ponseca Jr., Mark Sutton, Mercouri G. Kanatzidis and David G. Cooke

**Competing interests:**
Authors declare no competing interests.

**Data and materials availability:**
All data is available in the main text or the supplementary materials.

Supplementary Materials for

# Ultrafast correlated charge and lattice motion in a hybrid metal halide perovskite

Yang Lan[1], Benjamin J. Dringoli[1], David A. Valverde-Chavez[1], Carlito S. Ponseca Jr.[2], Mark Sutton[1], Yihui He[3], Mercouri G. Kanatzidis[3], & David G. Cooke[1]*

Correspondence to: cooke@physics.mcgill.ca

## Materials and Methods

### MAPI crystal synthesis

Single crystals of MAPI were grown from γ-butyrolactone (GBL) by the temperature-rising method with procedures similar to previous works (*31*). The binary precursors, PbI2 (5N, Sigma-Aldrich) and MAI (Sigma-Aldrich), were mixed with a stochiometric ratio of 1 to 1 in GBL solvent at 60-70 ∘C to form a 1.4 M solution. To facilitate the growth of single crystals, the as-prepared solution was then gradually heated to 100-110 °C. Once single crystals had nucleated, the solution temperature was kept constant to promote the continuous growth of large crystals. The as-grown single crystals were harvested after a period of several hours, yielding a cm$^3$-sized single crystal with large and flat facets suitable for THz spectroscopy.

### Transient multi-THz spectroscopy

Time-resolved THz spectroscopy (TRTS) measurements were performed using an all-air plasma based system providing single-cycle, coherently detected THz pulses continuously spanning the 1 -- 20 THz range. Broadband THz light was generated by a two-color laser plasma by focusing 800 nm, 35 fs pulses with an energy of 1 mJ and their second harmonic into dry air. No compensation for the group delay dispersion between 800 nm and 400 nm pulses was performed so the peak THz field strength is limited to ~30 kV/cm. This field strength, while sufficient to drive nonlinear polaron motion in a weakly coupled system such as GaAs, is not expected to significantly perturb the lattice potential for MAPI. A near-infrared pump pulse excites charge carriers at the band edge at a centre wavelength of ~795nm (1.56 eV), a fluence of 200 μJ/cm$^2$ and a pump-probe delay $\tau$. Optical pump - THz probe experiments were performed at normal incidence reflection from the $\langle 11\bar{1} \rangle$ facet of a large, ~1 cm$^3$-sized MAPI crystal of rhombic dodecahedral habit. Other facets showed similar results and are shown in the Supplementary Materials. Time and energy-resolved transient THz reflectivity maps were obtained by Fourier analysis of the reflected THz pulses without photoexcitation, $\tilde{E}_{ref}(\omega)$, and with, $\tilde{E}_{pump}(\omega,\tau)$, at a delay time $\tau$. Differential reflectivity is defined as $\frac{\Delta \tilde{r}(\omega,\tau)}{\tilde{r}_0(\omega)} = (\tilde{E}_{pump} - \tilde{E}_{ref})/\tilde{E}_{ref}$ and is complex-valued with full amplitude, shown in the main text, and phase information given in the Supplementary Materials. Measurements were performed at 295 K in the tetragonal phase of MAPI under dry purged gas conditions with no evidence of surface degradation during the ~20 hour map acquisition time.

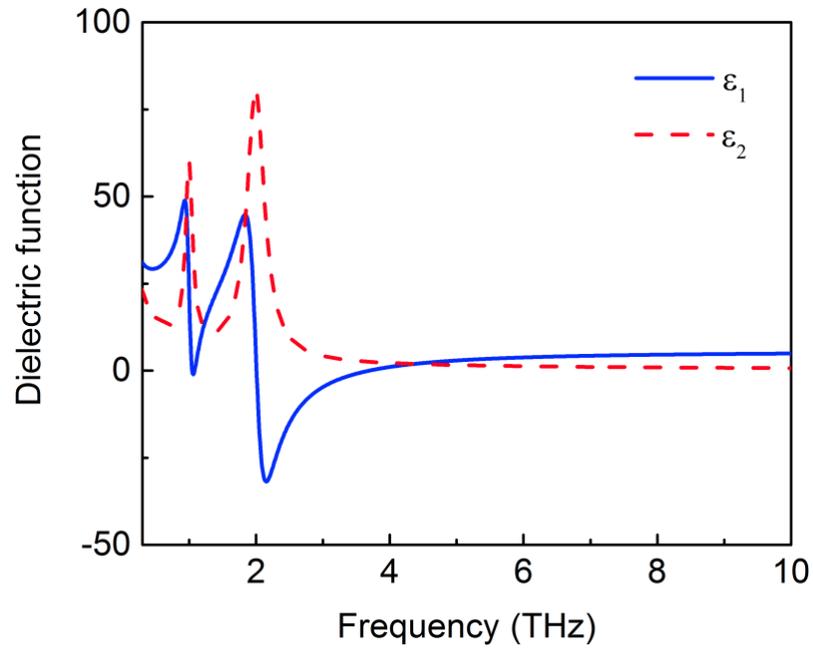

**Fig. S1. Static dielectric function of MAPI.** Real part (blue solid line) and imaginary part (red dash line) of MAPI dielectric function from the model in Ref (*28*).

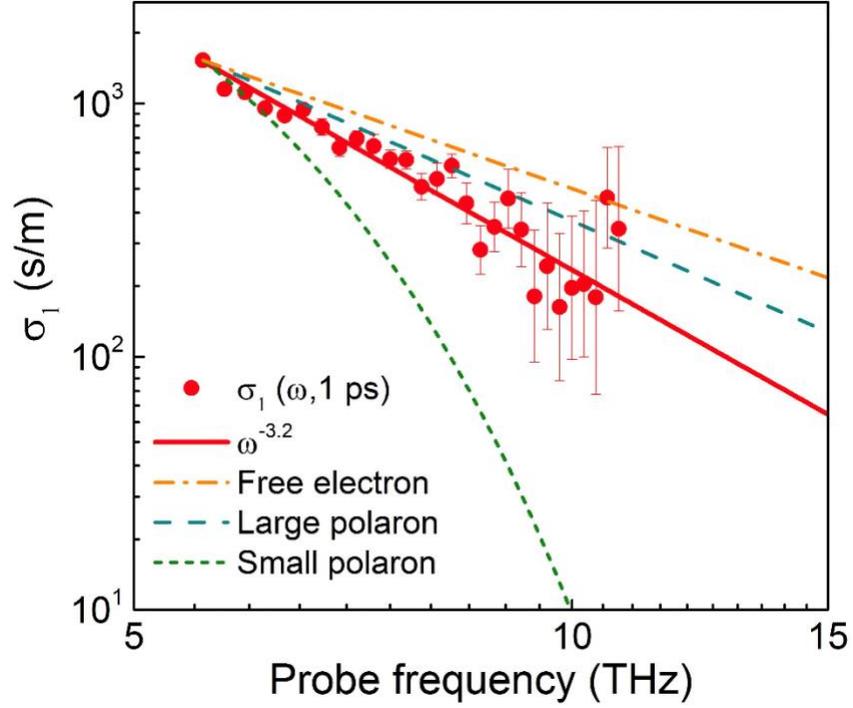

**Fig. S2. Conductivity of MAPI at $\tau$ =1 ps.** Real part of high frequency conductivity (red dots) with error bars representing standard deviation by THz reflection measurement at ⟨111⟩ facet, the power law fitting (solid line) with $\omega^{-3.2\pm0.2}$, the free electron case (dash dots) with $\omega^{-2}$, the large polaron limit (dash line) with $\omega^{-2.5}$, and the small polaron limit (short dash line) with a Gaussian shape rolling down.

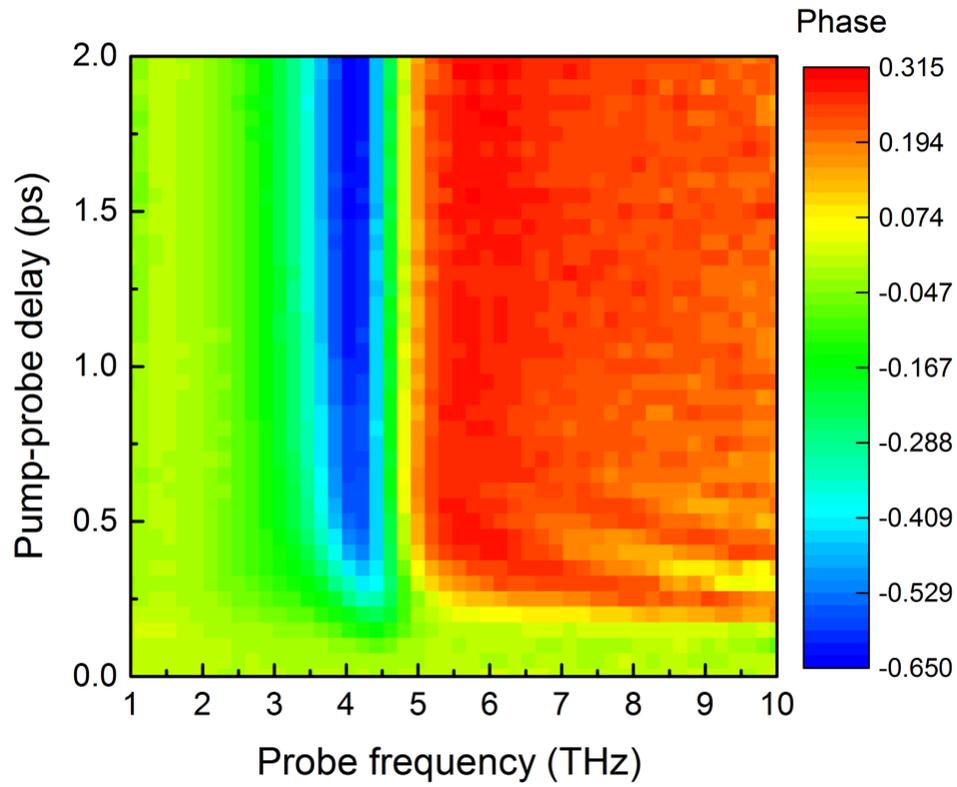

**Fig. S3. Phase of pump-induced differential THz reflectivity at the $\langle 11\bar{1} \rangle$ facet.** Two-dimensional pump-probe delay time/probe frequency map of phase for $\Delta r(\omega, \tau)/r_0$ at the $\langle 11\bar{1} \rangle$ facet.

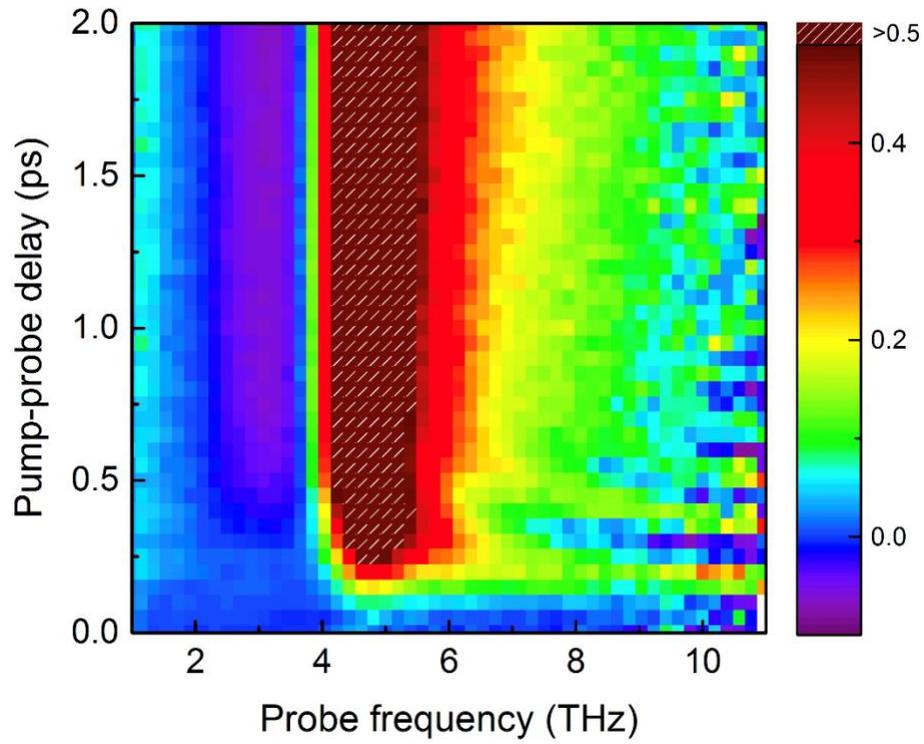

**Fig. S4. Amplitude of pump-induced differential THz reflectivity at the ⟨100⟩ facet.** Two-dimensional pump-probe delay time/probe frequency map of phase for $\Delta r(\omega,\tau)/r_0$ at the ⟨100⟩ facet under the same TRTS measurement at the ⟨10**0**⟩ facet.

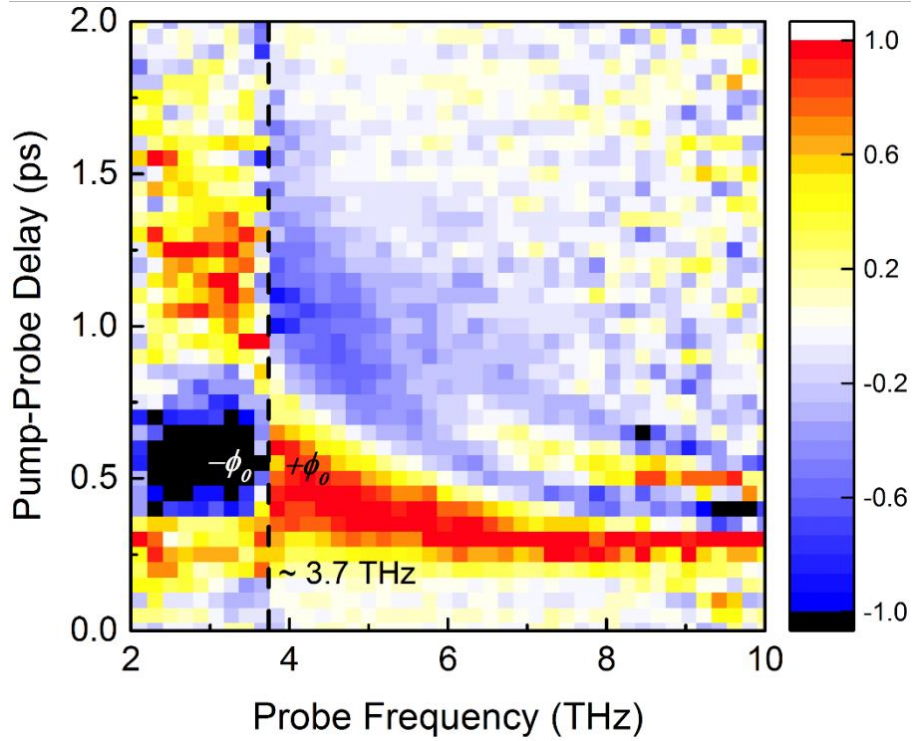

**Fig. S5. Time domain coherent oscillations of THz reflectivity without slow rise component at ⟨100⟩ facet.** Normalized $|\Delta r(\omega,\tau)|_{osc}^{norm}$ map with slow rise component subtracted. The vertical dash line marks the LO phonon frequency at 3.7 THz where a π phase flip occurs and above which a subsequent frequency chirp is evident.

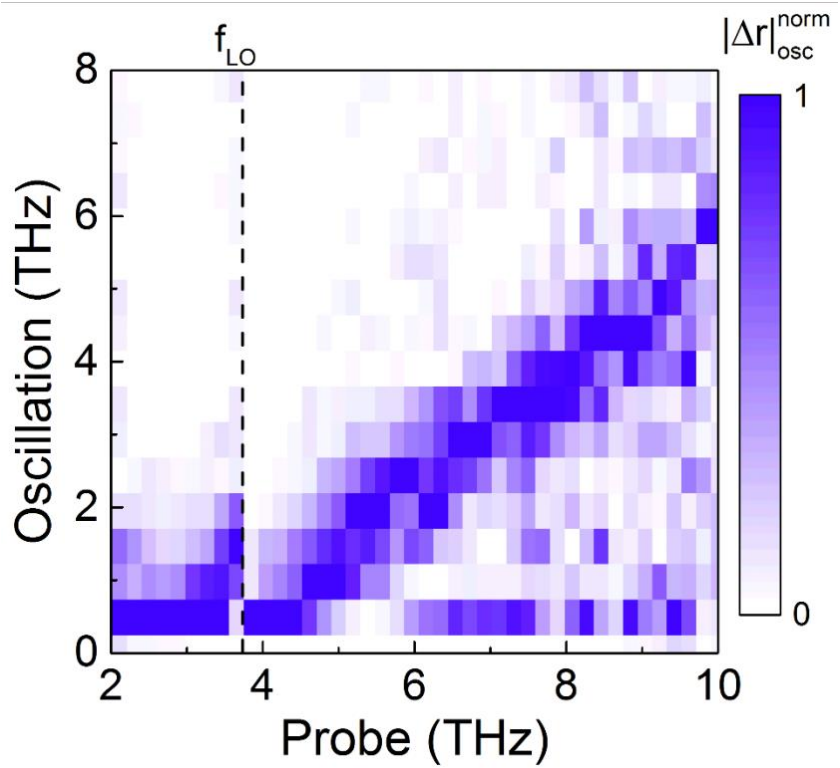

**Fig. S6. Fourier domain coherent oscillations of THz reflectivity without slow rise component at ⟨100⟩ facet.** Fourier transformation along delay time axis τ, $|\Delta r(\omega, \omega_{osc})|^{norm}_{osc}$, showing a clear onset of a coherent beat between charge motion at ω and $\omega_{LO}$. Information below 0.5 THz in oscillation is restricted by noise.

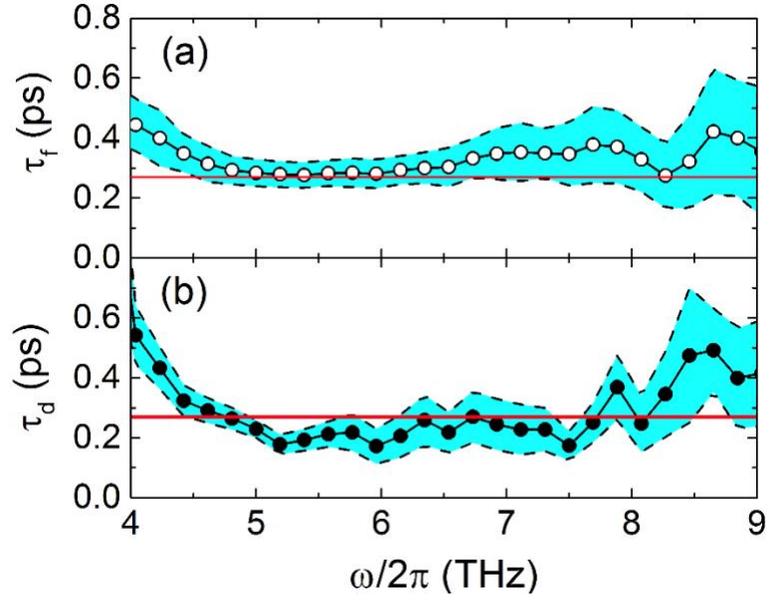

**Fig. S7. Time constants at ⟨100⟩ facet.** (a) Polaron formation time constant $\tau_f$ (open circles) with 95 % confidence bounds (shaded area). (b) Quantum beat dephasing time constant $\tau_d$ (closed circles) with 95 % confidence bounds (shaded area). The red line marks one phonon period.

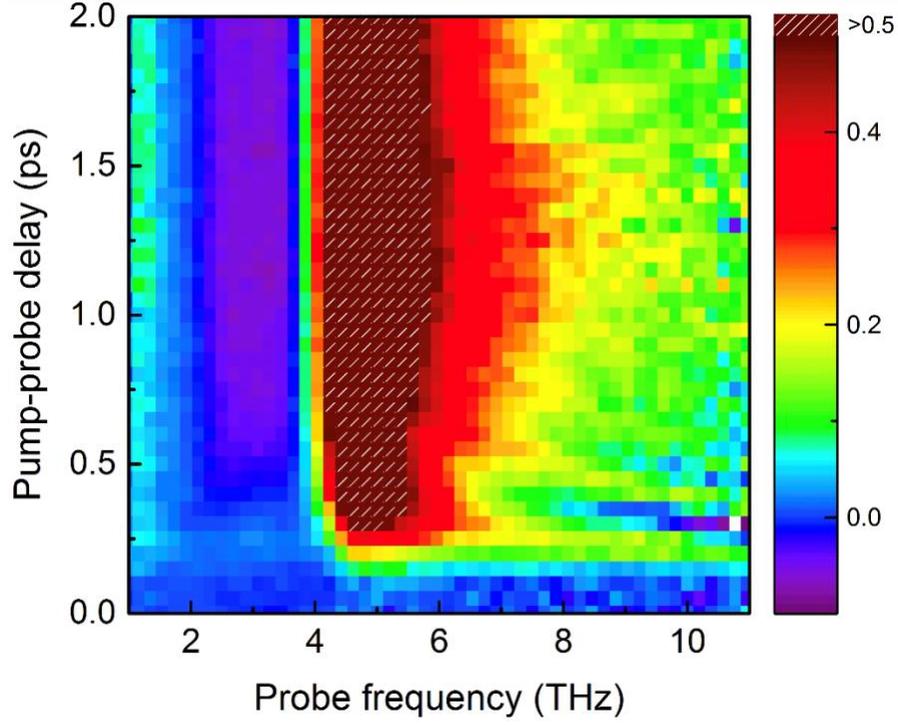

**Fig. S8. Amplitude of pump-induced differential THz reflectivity at ⟨111⟩ facet.** Two-dimensional time/energy map of amplitude for $\Delta r(\omega,\tau)/r_0$ for the ⟨111⟩ facet, under the same TRTS measurement for the ⟨11$\bar{1}$⟩ facet.

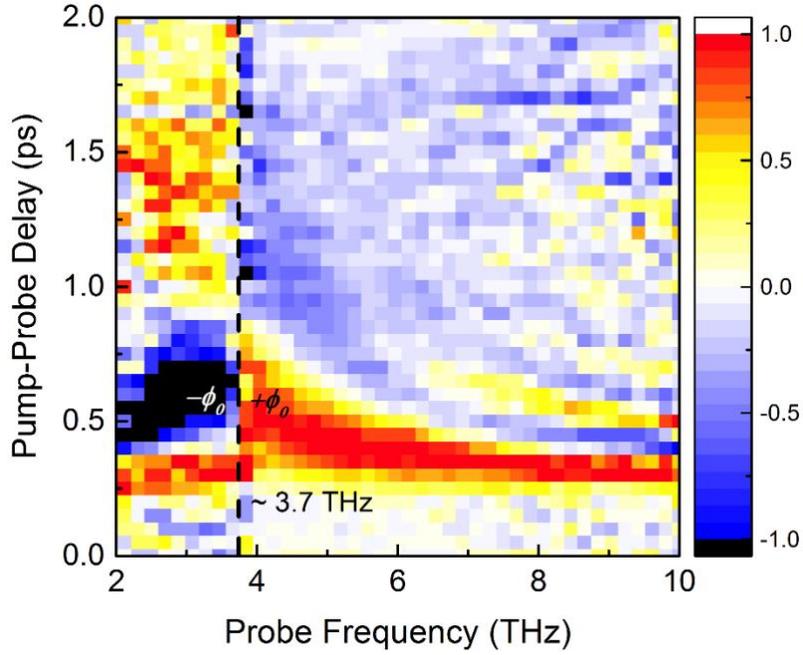

**Fig. S9. Time domain coherent oscillations of THz reflectivity without slow rise component at ⟨111⟩ facet.** Normalized $|\Delta r(\omega,\tau)|_{osc}^{norm}$ map with slow rise component subtracted. The vertical dash line marks the LO phonon frequency at 3.7 THz where a π phase flip occurs and above which a subsequent frequency chirp is evident.

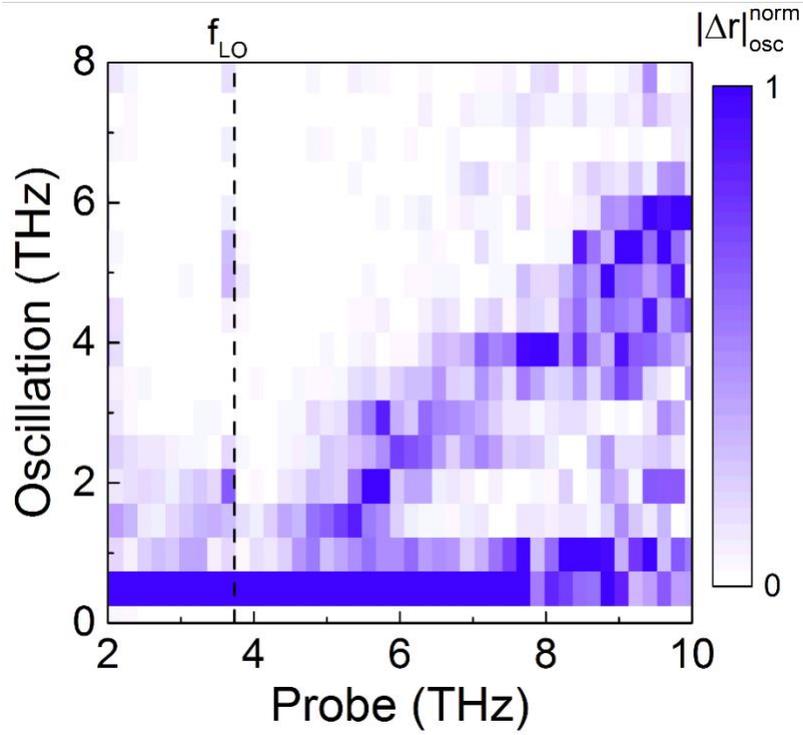

**Fig. S10. Fourier domain coherent oscillations of THz reflectivity without slow rise component at ⟨111⟩ facet.** Fourier transformation along delay time axis $\tau$, $|\Delta r(\omega, \omega_{osc})|^{norm}_{osc}$, showing a clear onset of a coherent beat between charge motion at $\omega$ and $\omega_{LO}$. Information below 0.5 THz in oscillation is restricted by noise.

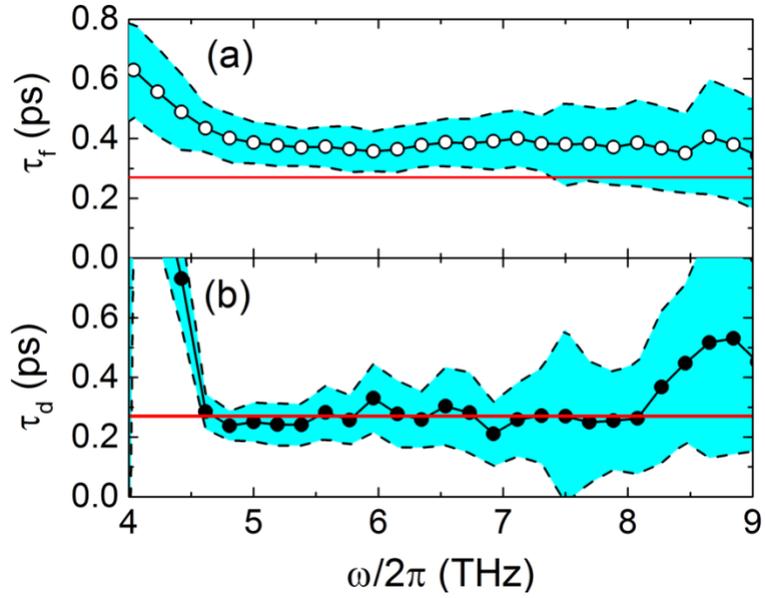

**Fig. S11. Time constants at ⟨111⟩ facet.** (a) Polaron formation time constant $\tau_f$ (open circles) with 95 % confidence bounds (shaded area). (b) Quantum beat dephasing time constant $\tau_d$ (closed circles) with 95 % confidence bounds (shaded area). The red line marks one phonon period.

| m* | ℏω$_{LO}$ | ε$_S$ | ε$_\infty$ |
|---|---|---|---|
| 0.23m$_0$ (*32*) | 15.3meV (*33*) | 30 (*28*) | 5.5 (*21*) |

**Table S1. Parameters describing MAPI dielectric properties.** Electron effective mass (m*), LO phonon frequency (ω$_{LO}$), static and optical dielectric constants (ε$_S$ and ε$_\infty$).